\documentclass[a4paper,aps,prl,floatfix,twocolumn,showpacs,superscriptaddress]{revtex4}
\usepackage[pdftex]{graphicx}
\usepackage{latexsym,amsmath,verbatim}
\usepackage{amsfonts,amsmath,amssymb,hyperref}
\usepackage{color}
\usepackage{hyperref}

\newcommand{\av}[1]{\langle {#1} \rangle}

\begin{document}

\title{Nature of the epidemic threshold for the susceptible-infected-susceptible dynamics in networks}

\author{Marian Bogu\~n\'a} 

\affiliation{Departament de F\'{\i}sica Fonamental, Universitat de
  Barcelona, Mart\'{\i} i Franqu\`{e}s 1, 08028 Barcelona, Spain}

\author{Claudio Castellano} 

\affiliation{Istituto dei Sistemi Complessi (ISC-CNR), via dei Taurini
  19, I-00185 Roma, Italy}

\affiliation{Dipartimento di Fisica, ``Sapienza'' Universit\`a di
  Roma, P.le A. Moro 2, I-00185 Roma, Italy}

\author{Romualdo Pastor-Satorras} 

\affiliation{Departament de F\'\i sica i Enginyeria Nuclear,
  Universitat Polit\`ecnica de Catalunya, Campus Nord B4, 08034
  Barcelona, Spain}

\date{\today}

\begin{abstract}
  We develop an analytical approach to the
  susceptible-infected-susceptible (SIS) epidemic model that allows us to
  unravel the true origin of the absence of an epidemic threshold in
  heterogeneous networks. We find that a delicate balance between the
  number of high degree nodes in the network and the topological
  distance between them dictates the existence or absence of such a
  threshold. In particular, small-world random networks with a degree
  distribution decaying slower than an exponential have a vanishing
  epidemic threshold in the thermodynamic limit.

\end{abstract}

\maketitle

The accurate theoretical understanding of epidemic thresholds on
complex networks is a pressing challenge in the field of network
science~\cite{barabasi02,mendesbook,Newman2010}. Indeed, such a
knowledge has potential practical applications in the design of
optimal immunization
programs~\cite{PhysRevE.65.036104,PhysRevLett.91.247901} and may shed
light on the behavior of the viral spreading of rumors, fads and
beliefs~\cite{goffman64,Leskovec:2007:DVM:1232722.1232727}.  In this
respect, a large research effort has been recently devoted to the
study of the susceptible-infected-susceptible (SIS) epidemic
model~\cite{anderson92}, the simplest model of epidemic spreading
showing an absorbing state phase transition between a healthy and an
endemic phase at a critical value of the effective infective rate
$\lambda$~\cite{marro1999npt}. The behavior of the
SIS model is particularly relevant in the case of highly heterogeneous
networks, for which a vanishing epidemic threshold in the
thermodynamic limit has been pointed out~\cite{pv01a}. Recently, a
scientific controversy has arisen concerning the location and the real
nature of the epidemic threshold in this kind of
networks~\cite{PhysRevLett.105.218701,Castellano12,PhysRevLett.109.128702,lee_epidemic_2012}.
In this paper, we provide strong analytical and numerical arguments 
showing that the threshold asymptotically
vanishes in any network with a degree
distribution decaying slower than exponentially, thus clarifying the
physical origin of this behavior.

In the SIS model, individuals can be in one of two states, 
either susceptible or infected. Susceptible individuals become infected
by contact with infected individuals at rate $\lambda$ times the number of 
infected contacts. Infected individuals, on the other hand, become 
spontaneously healthy again at rate $\mu$ that, without loss of 
generality, is set to unity.
The original approach to the dynamics of the SIS model~\cite{pv01a}
was based on the so-called heterogeneous mean-field (HMF)
theory~\cite{dorogovtsev07:_critic_phenom,barratbook}, which neglects
both dynamical and topological correlations. To do so, the actual
quenched structure of the network---given by its adjacency matrix
$A_{ij}$~\cite{Newman2010}---is replaced by an annealed version, in
which edges are constantly rewired at a rate much faster than that of
the epidemics, while preserving the degree distribution
$P(k)$. According to HMF theory, the epidemic threshold of the SIS
model takes the form $\lambda_c^\mathrm{HMF} = \av{k}/\av{k^2}$
\cite{pv01a}, where $\av{k}$ and $\av{k^2}$ are the first and second
moments of $P(k)$ \cite{Newman2010}. Many real networks have a
heterogeneous degree distribution, often scaling as a power-law (PL),
$P(k) \sim k^{-\gamma}$~\cite{barabasi02,mendesbook,Newman2010}. This
implies that the second moment diverges with the maximum degree
$k_{max}$ for a degree exponent $\gamma<3$, leading to a threshold
scaling $\lambda_c^\mathrm{HMF} \sim k_{max}^{\gamma-3}$, which
vanishes in the thermodynamic limit. On the other hand, for
$\gamma>3$, the second moment is finite and consequently so the
epidemic threshold.

While HMF theory represents an exact result in the case of annealed
networks~\cite{boguna09:_langev,Ferreira12}, its validity for real
(quenched) networks is limited. Indeed, an important improvement over
HMF theory is given by the quenched mean-field theory
(QMF)~\cite{1284681,VanMieghem09,Gomez_2010} which, while still neglecting
dynamical correlations, takes into account the full form of $A_{ij}$.
Within this framework, the epidemic threshold is predicted to be
$\lambda_c^\mathrm{QMF}= 1 / \Lambda_N$, where $\Lambda_N$ is the
largest eigenvalue of the adjacency matrix. Given the scaling of
$\Lambda_N$ with the maximum degree, $\Lambda_N \sim
\max\{\sqrt{k_{max}}, \av{k^2}/\av{k}\}$~\cite{Chung03}, QMF theory
predicts the same result as HMF theory for $\gamma<5/2$, while for
$\gamma>5/2$ it leads to $\lambda_c^\mathrm{QMF} \sim
1/\sqrt{k_{max}}$, that is, to a vanishing threshold for any value of
$\gamma$ (even for $\gamma>3$) in the thermodynamic
limit~\cite{PhysRevLett.105.218701}.  The prediction of QMF theory has
been validated for $\gamma<3$ by means of large scale numerical
simulations based on the quasi-stationary state
method~\cite{Ferreira12}. Numerical evidence for $\gamma>3$ is,
however, less convincing and has led to the following two criticisms.

Goltsev \textit{et al.}~\cite{PhysRevLett.109.128702} have considered,
within the QMF framework, the effects of eigenvector localization on
the steady state of the SIS model.  According to their observations,
in PL networks with $\gamma < 5/2$, the principal eigenvector is
delocalized, which implies that the density of infected nodes is
finite above $\lambda_c^\mathrm{QMF}$. However, for $\gamma>5/2$, the
principal eigenvector is localized, meaning that above
$\lambda_c^\mathrm{QMF}$ the system is active but activity is
concentrated around the hubs and their neighbors, leading to a number
of infected nodes that scales sub-linearly with system size and,
therefore, does not constitute a true endemic state. The endemic state
should, instead, appear at a different threshold, inversely
proportional to the eigenvalue of the upper delocalized state, and
approximately corresponding to the HMF
value~\cite{PhysRevLett.109.128702}. Therefore, the true threshold to
the endemic state would have a finite value for $\gamma>3$,
at odds
with the interpretation of QMF theory made in~\cite{PhysRevLett.105.218701}.

This view is further pursued by Lee \textit{et
  al.}~\cite{lee_epidemic_2012} by partly taking into account
dynamical correlations. Their argument is as follows: Slightly above
the QMF threshold, hubs in a PL network become active but their
activity is restricted to their immediate neighborhood. This activity has a characteristic lifetime
$\tau(k,\lambda)$ depending on the degree and the value of
the spreading rate $\lambda$. When hubs are directly connected to each
other (the case of a \textit{clustered} network, in the nomenclature
of Ref.~\cite{lee_epidemic_2012}), activity can be transferred between
hubs if the lifetime $\tau(k,\lambda)$ is sufficiently large. In this
case, above $\lambda_c^\mathrm{QMF}$ the network is able to support an
endemic state, characterized by the mutual reinfection of
connected hubs. In the case of unclustered networks, however, hubs are
not directly connected and the reinfection mechanism does not
work. Thus, the authors of Ref.~\cite{lee_epidemic_2012} claim that
the state above $\lambda_c^\mathrm{QMF}$ is just a Griffiths phase
\cite{Vojta}, where the density of infected nodes decays with time
more slowly than exponentially (logarithmically indeed), while the
actual epidemic threshold is located at a higher, finite value of
$\lambda$. Within this picture a true zero epidemic threshold in the
thermodynamic limit occurs only for $\gamma <3$.

While the arguments presented in
Refs.~\cite{PhysRevLett.109.128702,lee_epidemic_2012} are appealing
and, apparently, leading to the conclusion that the threshold is
finite in random PL networks with $\gamma>3$, here we reconsider the
problem and provide analytical and numerical evidence pointing in the opposite
direction, namely, \textit{a vanishing epidemic threshold for any
  small-world network with a degree distribution decaying slower than
  exponentially, in particular power-law networks with any
  $\gamma$}. To confirm this prediction, we propose a numerical
approach, based on the scaling analysis of the survival time of the
infection process, which is able to provide very accurate estimates of
the epidemic threshold even in the regime where the quasi-stationary
state method is unreliable.

Our analytical approach is based on the consideration of dynamical
correlations, as in~\cite{lee_epidemic_2012}, but not
restricted to
direct neighbors.  The argument of~\cite{lee_epidemic_2012} assumes
that a zero epidemic threshold can only occur in clustered
networks, when hubs are directly connected to each other and can
reinfect each other within a time smaller that the characteristic
healing time $\tau$.  However, as already pointed out in
Ref.~\cite{Chatterjee09}, a direct connection is not a necessary
condition for the possibility of hub reinfection. Instead, we should
properly consider the possibility of reinfection between two vertices
$i$ and $j$, separated by a topological distance $d_{ij}$, possibly
larger than $1$. Indeed, the epidemic threshold predicted by the HMF
theory is actually based on the local properties of the network alone,
assuming that the local structure will replicate in a tree like
fashion forever, preserving only the statistical properties of the
network. Then, above $\lambda_c^\mathrm{HMF}$, we expect that a
perturbation originated in a node will be able to propagate as a
supercritical branching process forever. Below this threshold, this
process is not possible.  Still, as we show below, the epidemics can
sustain itself, due to perturbations which propagate up to distances
of order $\ln(N)$, where $N$ is the network size.

To take into account dynamical correlations over distant neighbors, 
we replace the original SIS dynamics by a modified description of 
the SIS process valid over coarse-grained time scales. On such longer
temporal intervals, it is possible that a given infected node $i$
propagates the 
the infection to any other node $j$ in the network via a
sequence of microscopic infection events of intermediate, nearest
neighbors  nodes.
The infective rate $\lambda$ is then replaced by the effective
rate $\bar{\lambda}(d_{ij},\lambda)$ at which the infected node
$i$ infects any other node $j$ in the network
when the process is mediated by a chain of $d_{ij}-1$
intermediate nodes. 
On the coarse-grained time scale also the recovery rate $\delta$
of node $i$  is replaced by an effective rate $\bar{\delta}(k_i,\lambda)$.
Overall, the evolution of the SIS dynamics over the coarse-grained
time scale is then given by
\begin{equation}
  \frac{d \rho_i(t)}{dt}=-\bar{\delta}(k_i,\lambda) \rho_i(t)+
  \sum_{j \neq i} \bar{\lambda}(d_{ij},\lambda)
  \rho_{j}(t)[1-\rho_i(t)],
  \label{eq:1}
\end{equation}
which is defined on a fully connected graph.  The parameters
$\bar{\delta}$ and $\bar{\lambda}$ reflect in this description the
structure of the original network.  On long time scales node $i$ is
considered as susceptible only when the node and all of its nearest
neighbors in the original graph are susceptible: hence its recovery
rate is $\bar{\delta}(k_i,\lambda)=\tau^{-1}(k_i,\lambda) \approx
e^{-a(\lambda) k_i}$ (see SI for numerical results and an analytical
argument \cite{supplement}), where $a(\lambda)$ is a smooth growing
function of $\lambda$.  To evaluate the effective infective rate it is
convenient to assume that paths connecting nodes are independent and
made of nodes of degree $2$. This leads to the expression
$\bar{\lambda}(d_{ij},\lambda) \approx \lambda
e^{-b(\lambda)(d_{ij}-1)}$, with $b(\lambda) = \ln(1+1/\lambda)$ (see
SI for an analytical derivation and a numerical
validation~\cite{supplement}).  The assumptions made for determining
$\bar{\lambda}$ are clearly not true in a real network because there
are many paths connecting the same pair of nodes and intermediate
nodes have, in general, degrees larger than 2.  This implies that the
infective rate between two nodes that we use in the coarse-grained SIS
dynamics is smaller than the real one.  Therefore, Eq.~(\ref{eq:1})
will provide an upper bound for the true epidemic threshold of the
original SIS dynamics and, thus, the absence of an epidemic threshold
of the former will imply also its absence in the latter.

Eq.~(\ref{eq:1}) can be applied to any network. We can, however, get
deeper insights in the case of small-world random graphs, in which the
the average internode topological distance takes the
form~\cite{PhysRevE.72.026108}
\begin{equation}
  d_{kk'}=1+\frac{\ln{\left(\frac{N \langle k \rangle}{k
          k'}\right)}}{\ln{\kappa}} 
\label{eq:2}
\end{equation}
where $\kappa = \av{k^2} / \av{k} - 1$ is the average branching factor
of the network. Using the approximation Eq.~(\ref{eq:2}) allows us to
coarse grain Eq.~(\ref{eq:1}) for degree classes. After defining
$\rho_k(t)\equiv \sum_{deg(i) = k} \rho_i(t)/NP(k)$ and plugging
Eq.~(\ref{eq:2}) into Eq.~(\ref{eq:1}), we obtain
\begin{eqnarray}
  \lefteqn{\frac{d \rho_k(t)}{dt}=-\bar{\delta}(k,\lambda) \rho_k(t)}
\\
 &+&\lambda N \left[ \frac{k}{N \langle k
     \rangle}\right]^{\frac{b(\lambda)}{\ln{\kappa}}} 
  \sum_{k'} k'^{\frac{b(\lambda)}{\ln{\kappa}}}P(k') 
  \rho_{k'}(t)[1-\rho_k(t)]. \nonumber
\label{eq:3} 
\end{eqnarray}
Notice that the use of Eq.~(\ref{eq:2}) implies that the local
propagation of the infection among directly connected nodes is
neglected; only reinfections between distant ($\sim \ln N$) nodes are
taken into account.  By performing a linear stability analysis of this
equation, we can see that the critical epidemic threshold of the 
coarse-grained SIS dynamics, $\lambda_c$, is the solution of the
transcendental equation (see SI for a detailed derivation
\cite{supplement})
\begin{equation}
1=\lambda N \sum_{k=k_{min}}^{k_{max}} P(k) \tau(k,\lambda) \left[
  \frac{k^2}{N \langle k
    \rangle}\right]^{\frac{b(\lambda)}{\ln{\kappa}}}. 
\label{eq:lambdac}
\end{equation}
In general, the maximum degree of the network, $k_{max}$, is a growing
function of $N$. If we further assume that the degree distribution
decays slower than an exponential, Eq.~(\ref{eq:lambdac}) can be
approximated as the integral near the upper bound $k_{max}$, i.e.
\begin{equation}
1=\frac{\lambda}{a(\lambda)}
 e^{a(\lambda) k_{max}-\frac{b(\lambda)}{\ln{\kappa}}\ln\left[ \frac{N
       \langle k
       \rangle}{k_{max}^2}\right]-\ln\left[\frac{1}{NP(k_{max})}
   \right]}. 
\end{equation}
When $P(k)$ decays slower than an exponential, $k_{max}$ grows faster
than $\ln{N}$. Therefore, as the system size grows while keeping
$\lambda$ fixed, there is a point where the first term in the
exponential becomes larger than the other two (negative) terms and,
eventually, the right hand side of Eq.~(\ref{eq:lambdac}) becomes
larger than $1$. {\it As a consequence, the epidemic threshold of the
 coarse-grained SIS dynamics starts decreasing as $N$
  increases, thus going
  to zero in the thermodynamic limit}. Making the additional
assumption that $a(\lambda) \approx a \lambda^2$ for $\lambda \ll 1$
(which is compatible with numerical simulations, see 
SI~\cite{supplement}), we conclude that the upper bound of the epidemic
threshold decreases as $1/\sqrt{k_{max}}$, with additional logarithmic
corrections to scaling. Interestingly, this scaling is similar to the
one predicted by the QMF theory. However, in our case, the threshold
marks the onset of a true endemic state where a finite fraction of all
nodes of the system are active.

The case of non small-world networks can be considered along the same
lines. Unfortunately, a general formula for the average topological
distance as a function of nodes' degrees is not known. Nevertheless,
the absence of long range connections in non small-world networks
suggests that node degree is not as determinant as in the case of
small-world ones. Thus, to get some understanding, we assume an
internode distance independent of the degree and scaling as a
power-law with system size, i.e. $d = 1 + \alpha N^\beta$. An analysis
similar to the case of small-world networks (see SI \cite{supplement})
concludes that {\it non small-world networks have a vanishing epidemic
  threshold only if $k_{max}$ grows faster than $N^{\beta}$}.  This
result explains the finite epidemic threshold in the $(3,3)$-flower
model~\cite{rozenfeld_fractal_2007} found in Ref.~\cite{lee_epidemic_2012}, even if the model
generates a PL network with $\gamma=1+\ln{6}/\ln{2} \approx
3.58$. Indeed, this model generates a non small-world network with
$\beta=\ln{3}/\ln{6}$ whereas $k_{max} \sim
N^{\ln{2}/\ln{6}}$~\cite{rozenfeld_fractal_2007}.

\begin{figure}[t]
\includegraphics*[width=1.\linewidth]{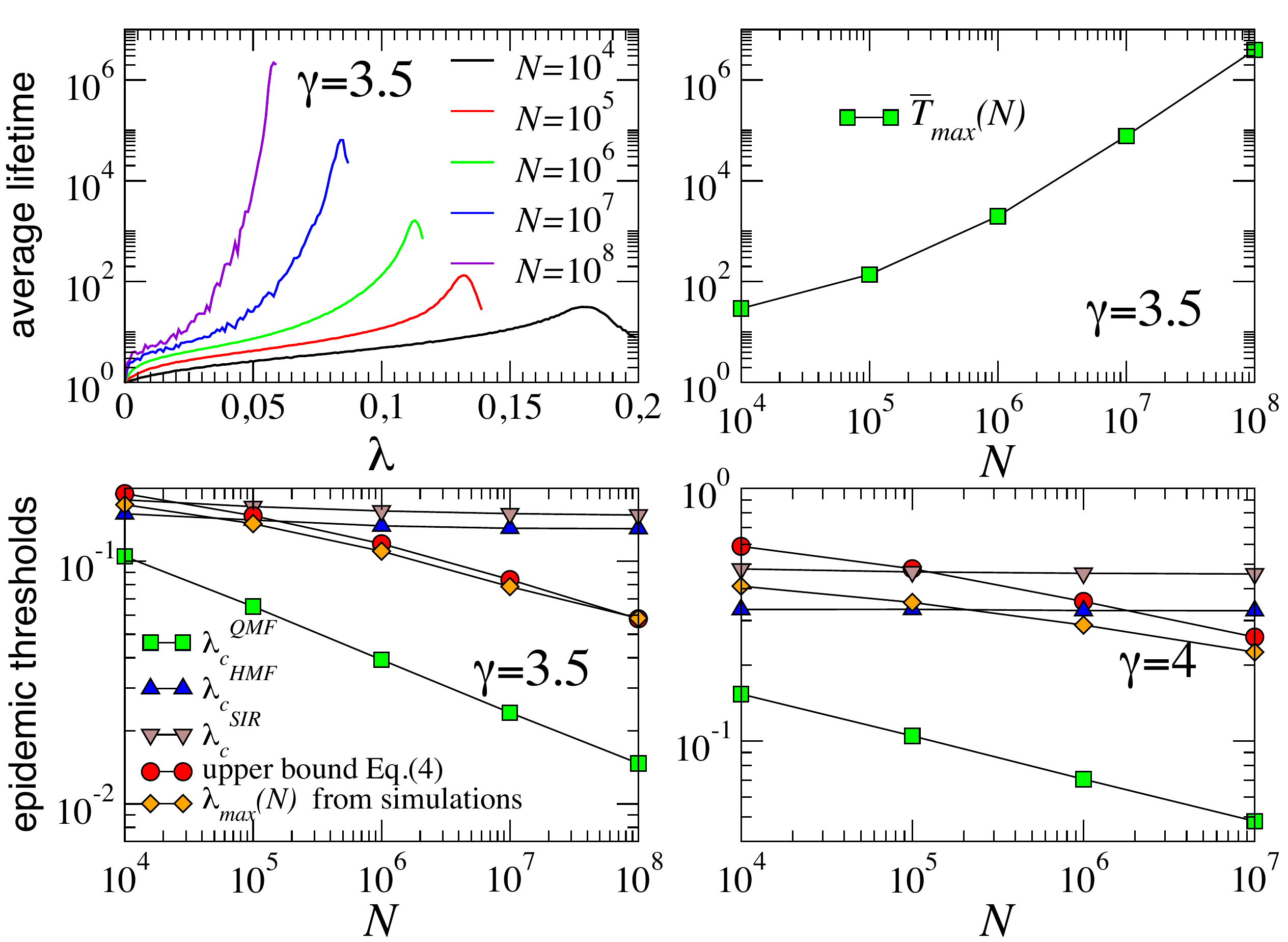}
\caption{(color online) Top plots (single network realization): (left)
  Average lifetime of finite
  realizations for PL random graphs with $\gamma=3.5$, $k_{min}=3$ 
  and different sizes; 
  (right) divergence of the height of the
  peak in $\bar{T}(\lambda)$ as a function of the network size. Bottom
  plots (averages over 5 network realizations): 
  Epidemic thresholds as measured in numerical simulations in
  PL networks of different sizes, $\gamma=3.5$, $k_{min}=3$ and $\gamma=4$,
  $k_{min}=2$. It is also shown the upper bound predicted by our
  theory as found from a numerical solution of Eq.~(\ref{eq:lambdac})
  with $\tau(k,\lambda)$ measured in numerical simulations. For the
  sake of comparison, we also show the QMF, HMF, and SIR thresholds.}
\label{fig:1} 
\end{figure}

To check the accuracy of our theory, we propose a method to estimate
the critical point of absorbing state phase transitions.  The method
is based on the analysis of individual realizations of the process
starting with a single infected node. Each realization is
characterized by its lifetime $T$ and coverage $C$, where the latter
is defined as the fraction of distinct nodes ever infected during the
realization.  In the thermodynamic limit, realizations can be of two
types: finite or endemic. Finite realizations have a finite lifetime
and, therefore, a vanishing coverage in the thermodynamic
limit. Endemic realizations, on the other hand, have an infinite
lifetime and their coverage is equal to 1. Below the epidemic
threshold, all realizations are trivially finite. Above the threshold,
there is a non null probability, $P_{end}(\lambda)$, that a
realization that starts at a single node becomes endemic, making
$P_{end}$ a good order parameter of the phase transition.  Akin to the
role of the average size of finite clusters in standard percolation~\cite{stauffer94}, 
in our approach the role of susceptibility is
played by the average lifetime of finite realizations
$\bar{T}(\lambda)$, which diverges at $\lambda_c$ both from below and
above.

In finite systems, the major problem is to determine when a
realization is endemic or not. One possibility is to declare a
realization as endemic whenever its coverage reaches 1.  However, from
a computational point of view, this option is too costly.  We
therefore take advantage of the following fact: In an infinite size
system, whenever the coverage of a realization reaches a finite
fraction (even small), the probability of the realization being
endemic is 1.  Then, in finite systems, we declare a realization as
endemic whenever its coverage reaches a predefined value (in our case
$C=0.5$, see SI for tests with other values~\cite{supplement}) and
stop the realization at this point.  Then, we can measure the average
lifetime of finite realizations $\bar{T}(\lambda,N)$ and the position
of its peak, which we take as the estimate of the epidemic threshold
for finite systems. Finally, we note that the method can be applied
starting from any node of the network with identical results as far as
the position of the threshold is concerned. Here, to minimize the
fluctuations of $\bar{T}(\lambda,N)$ close to the critical point, we
start our simulations always from the node with highest degree.

Figure~\ref{fig:1} shows the result of this program in random PL
networks generated with
the uncorrelated configuration model~\cite{ucmmodel}. The average
lifetime $\bar{T}(\lambda,N)$ behaves as an effective susceptibility
and, thus, we estimate the epidemic threshold for a finite network as
the position of its peak, $\lambda_{max}(N)$. These estimates are
shown in the bottom plots and compared with the upper bound given by a
numerical solution of Eq.~(\ref{eq:lambdac}) and where
$\tau(k,\lambda)$ is obtained from numerical simulations. As it can be
clearly seen, the upper bound predicted by our theory is in very good
agreement with numerical simulations, even for $\gamma=4$, a network
clearly ``unclustered'' according to~\cite{lee_epidemic_2012}.

Notice that, due to the approximation made in Eq.~(\ref{eq:2}), our
theory neglects the propagation of the epidemic mediated only by
connected nodes, which is the approach taken in the HMF
theory. Therefore, one should expect that the true upper bound for the
real epidemic threshold is the minimum between the estimation given by
Eq.~(\ref{eq:lambdac}) and $\lambda_c^{\mathrm{HMF}}$. From this
perspective, it is surprising that the epidemic threshold measured
from simulations is higher than $\lambda_c^{\mathrm{HMF}}$ for small
system sizes.  Notice, however, that the HMF theory of the SIS
dynamics completely neglects dynamical correlations. These
correlations account for the fact that, whenever a node is infected,
there is a high probability for the node that infected it to be still
infected.  Therefore, the number of neighbors available to an infected
node to further propagate the epidemics is, in most cases, its degree
minus 1.  Consequently, a better upper bound for the local propagation
of the dynamics is given by the HMF theory of the SIR model, that is
$\lambda_c^{SIR}=\langle k \rangle/\langle k(k-1) \rangle$. Bottom
plots of Fig.~\ref{fig:1} show the estimation of $\lambda_c^{SIR}$,
which is always above the real threshold.

To sum up, the behavior of the SIS epidemic threshold in networks depends
on a delicate balance between their local and global properties. Both
HMF and QMF theories are constructed by considering only the local
dynamics of infections among nearest neighbors, and thus fail to
provide a correct description. Here we have presented a theoretical
approach to epidemics in networks, built upon previously sketched
concepts, that takes into account the full network structure, and that
considers reinfection events among nodes not directly connected,
i.e. mediated by chains of other nodes. Our theoretical analysis,
while based in some (reasonable) approximations, is well backed up by
means of reliable numerical evidence. The main conclusion of both
approaches is that the epidemic threshold in SIS model is effectively
null in the thermodynamic limit in all random small-world networks
with a degree distribution decaying slower than exponentially. 
Beyond this remarkable result, our work highlights
the subtle role that dynamical correlations might play
in non-equilibrium heterogeneous systems near criticality.

\begin{acknowledgments}
  M.~B. acknowledges financial support from the Spanish MICINN project
  No.\ FIS2010-21781-C02-02; Generalitat de Catalunya grant No.\
  2009SGR838; and by the ICREA Academia prize, funded by the
  Generalitat de Catalunya. RPS acknowledges financial support from
  the Spanish MICINN, under project FIS2010-21781-C02-01 and
  additional support through ICREA Academia, funded by the Generalitat
  de Catalunya.
\end{acknowledgments}

\section{Supplementary Information}
\appendix

\section{Numerical simulations}

The SIS dynamics is simulated with a continuous time dynamics as
follows: During the course of the simulation, we keep track of the
number of infected nodes $N_I(t)$ and the number of active links
$E_A(t)$, where an active link is defined as a link emanating from an
infected node (notice that links connecting two infected nodes will
appear twice in this list). At each step, with probability
$p_r=N_I(t)/[N_I(t)+\lambda E_A(t)]$, a randomly chosen infected node
is turned susceptible whereas, with probability $1-p_r$, an active
link is chosen at random and if one of the two nodes attached to the
link is susceptible, then this node is turned infected. After this
procedure, time is updated as $t \rightarrow t+1/[N_I(t)+\lambda
E_A(t)]$ and the list of infected nodes and active links
recomputed. An equivalent algorithm keeps a list of active links as
those connecting one infected node and one susceptible. The advantage
of this latter method is that infectious attempts always end up with a
susceptible node being infected, which is not the case with the first
method. In this work, all simulations are backed up independently with
the two methods.

\section{Estimation of the infective rate $\bar{\lambda}(d_{ij},\lambda)$}

\begin{figure}[t]
\includegraphics*[width=\linewidth]{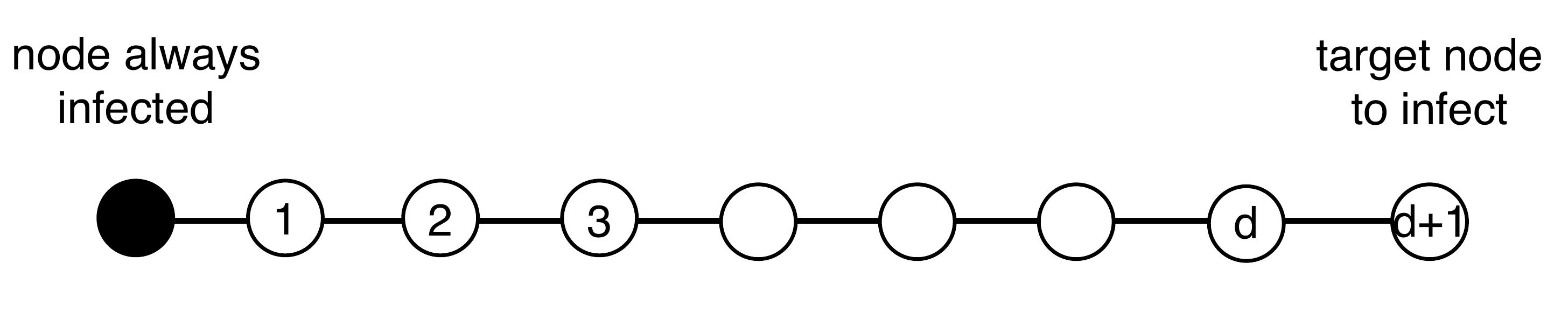}
\caption{Scheme of the infectious events mediated by chains of nodes.}
\label{figS1} 
\end{figure}

This rate is the inverse of the average time infected node $i$ takes
to infect node $j$ when they are separated by a distance $d_{ij}$ in
the original graph.  We consider the process of transmission of the
infection between the two nodes mediated by a one dimensional chain of
length $d_{ij}$. Consider a one dimensional chain of $d+2$ nodes with
the leftmost node always infected, as indicated in
Fig.~\ref{figS1}. Let $T(d+1)$ the average time the node at the
rightmost position takes to get infected for the first time. Let
$\tilde{T}(d)$ the average time between two consecutive infectious
events (after the first one) of the node at distance $d$. Because we
are in a chain and the source of the infection is the leftmost node,
the node at distance $d+1$ can only get infected for the first time by
its left neighbor. Once this node is infected, the probability that
the target node gets the infection before its left neighbor recovers
is simply given by
\begin{equation}
  p=\frac{\lambda}{1+\lambda}
\end{equation}
the node at distance $d+1$ can get infected right after its left node
gets infected for the first time or after the second time, and so
on. The probability that the node gets infected right after its left
neighbor gets infected for the $n-$th time is
\begin{equation}
\mbox{Prob}(n)=p(1-p)^{n-1}
\end{equation}
On the other hand, the average time elapsed in this case $T_n(d+1)$ is
\begin{equation}
T_n(d+1)=T_n(d)+(n-1)(1+\tilde{T}(d))+\lambda^{-1}.
\end{equation}
Combining these two results, we get the equation for the average
infection time, $T(d)=\sum_n \mbox{Prob}(n) T_n(d)$,
\begin{equation}
T(d+1)=T(d)+\frac{2}{\lambda}+\frac{1}{\lambda} \tilde{T}(d)
\end{equation}
with the initial conditions $T(1)=\tilde{T}(1)=\lambda^{-1}$. In the
limit of low infectious rate, that is, $\lambda \ll 1$, we can
approximate $\tilde{T}(d) \approx T(d)$ and we get a closed recursive
equation for $T(d)$, whose solution is
\begin{equation}
  T(d)=\frac{1}{\lambda} \left[ \left(1+ 2 \lambda \right) \left(1+
      \frac{1}{\lambda} \right)^{d-1} -2 \lambda\right] \approx
  \frac{1}{\lambda} e^{(d-1)\ln{\left(1+\frac{1}{\lambda}\right)}} 
\end{equation}
We then conclude that the infective rate is
\begin{equation}
  \bar{\lambda}(d_{ij},\lambda) \approx \lambda
  e^{-b(\lambda)(d_{ij}-1)}, \mbox{ with }
  b(\lambda)=\ln{\left(1+\frac{1}{\lambda}\right)} 
\end{equation}

In the case of small-world random graphs, the average internode
topological distance depends only on the degree of the nodes as
\cite{PhysRevE.72.026108}
\begin{equation}
  d_{k,k'}=1+\frac{\ln{\left(\frac{N \langle k \rangle}{k
          k'}\right)}}{\ln{\kappa}}.
\label{eq:2}
\end{equation}
Inserting this expression into the effective infective rate we get
\begin{equation}
\lambda_{k,k'} = \lambda \left[ \frac{k k'}
{\av{k}N}\right]^{\frac{b(\lambda)}{\ln{\kappa}}}.
\label{lambdakk}
\end{equation}

\begin{figure}[t]
\includegraphics*[width=\linewidth]{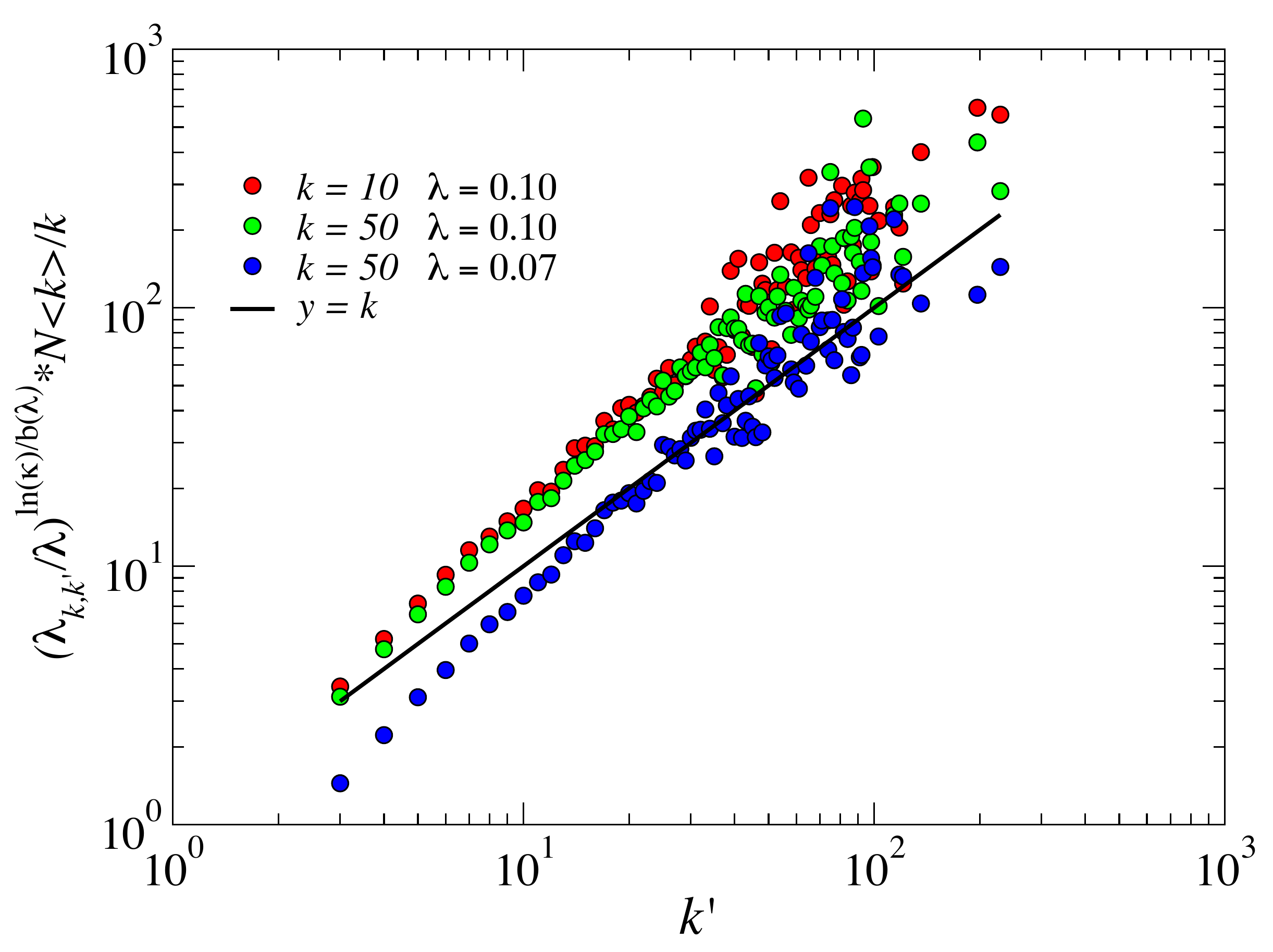}
\caption{Plot of $\left[\lambda_{k,k'}/\lambda\right]^{\ln{\kappa}/b(\lambda)}
\cdot N \av{k}/k$ vs $k'$ for various values of $\lambda$ and $k$.
The prediction of Eq.~(\ref{lambdakk}) is the straight solid line.}
\label{figS2} 
\end{figure}

We test numerically this relationship by keeping a single node of
degree $k$ always infected and computing the time 
it takes to infect for the first time any other node in the network.
The rate $\lambda_{k,k'}$ is obtained by inverting the value of this time for
the first infectious event, averaged over all nodes of degree $k'$.
Equation~(\ref{lambdakk}) predicts that plotting 
$\left[\lambda_{k,k'}/\lambda\right]^{\ln{\kappa}/b(\lambda)}$ vs $k'$
a linear behavior must be found, and this turns out to agree with
the outcome of simulations in a network with $\gamma=3.5$ (see Fig.~\ref{figS2}).

\section{Estimation of the recovery rate $\delta(k,\lambda)$}

\begin{figure}[t]
\includegraphics*[width=\linewidth]{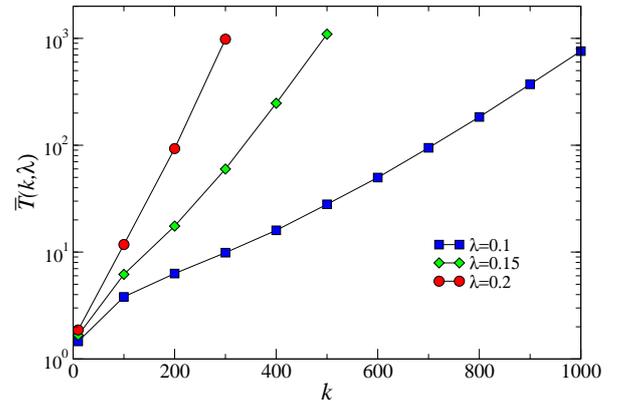}
\caption{Average lifetime of star graphs with the hub originally
  infected as a function of the star degree and different values of
  $\lambda$.}
\label{figS3} 
\end{figure}

This rate can be estimated as the inverse of the survival time of an
infection starting at the center of a star of degree
$k$. Unfortunately, the exact solution to this problem is rather
involved (see~\cite{Cator:2013lp} for an exact treatment). Here, we
present an approximation based on the discretization of the process in
time units of $\mu^{-1}=1$. Consider the following cycle: initially,
the center of the star --the hub-- is infected whereas leaf nodes are
susceptible. The probability that a leaf node is infected when the hub
recovers is
\begin{equation}
  p_{in}=\frac{\lambda}{2 +\lambda}.
\end{equation}
Then, by the time the hub recovers, there are $n$ infected leaf nodes
with probability
\begin{equation}
\mbox{Prob}(n|k)=\left(
\begin{array}{c}
k\\
n
\end{array}
\right)p_{in}^n(1-p_{in})^{k-n}
\end{equation}
the probability that at least one of these $n$ infected nodes infects
the hub again before they recover is
\begin{equation}
  \sum_{n=1}^k \mbox{Prob}(n|k)[1-(1-p_{in})^n]=1-(1-p_{in}^2)^k.
\end{equation}
The average time to complete the cycle is $2$.  The probability that
the outbreak goes through a sequence of $m$ complete cycles and then
dies is
\begin{equation}
(1-p_{in})^k \left[ 1-(1-p_{in}^2)^k \right]^m ; m=0,1,\cdots
\end{equation}
and the time elapsed $(2m+1)$.  However, an outbreak can also die in
the middle of the cycle, that is, when infected leaves recover before
infecting the hub again. The probability that the outbreak goes
through a sequence of $m$ complete cycles and dies in the middle of
the $m+1$ cycle is
\begin{equation}
\left[(1-p_{in}^2)^k-(1-p_{in})^k \right] \left[ 1-(1-p_{in}^2)^k \right]^m ; m=0,1,\cdots
\end{equation}
The average elapsed time is in this case $(2m+2)$.  Putting these
pieces together, the effective recovery rate can be approximated as
\begin{equation}
\delta^{-1}(k,\lambda)=\frac{2}{(1-p_{in}^2)^k}-\frac{1}{(1+p_{in})^k}.
\end{equation} 
For low infectious rates $p_{in} \ll 1$, this can be approximated as
\begin{equation}
\delta^{-1}(k,\lambda) \approx 2 e^{-\lambda^2 k/4}.
\label{deltafinal}
\end{equation}
As we have mentioned at the beginning of this section, the previous
calculations provide only an approximation to the true recovery
rate. This is so because we have considered the process as discretized
in time whereas the real process takes place at continuous
time. Nevertheless, we expect that Eq.~(\ref{deltafinal}) captures the
correct functional dependence. To check this result, we have performed
simulations of the SIS model on star graphs, starting from a state
with only the hub infected and computed the average time needed to
reach the absorbing healthy state.  Fig.~\ref{figS3} shows the average
lifetime for fixed values of $\lambda$ as a function of the degree
$k$, where the exponential trend predicted by our calculations is
clearly visible.

\section{Derivation of Eq.~(4)}

Let us consider Eq.~(3) in the main paper, namely
\begin{widetext}
\begin{equation}
  \frac{d \rho_k(t)}{dt}=-\bar{\delta}(k,\lambda) \rho_k(t)+
\lambda N \left[ \frac{k}{N \langle k
    \rangle}\right]^{\frac{b(\lambda)}{\ln{\kappa}}} 
  \sum_{k'} k'^{\frac{b(\lambda)}{\ln{\kappa}}}P(k') 
  \rho_{k'}(t)[1-\rho_k(t)]. 
  \label{eq:3}
\end{equation}
\end{widetext}
It is obvious that the absorbing state $\rho_k(t)=0$ is a fixed point
of the dynamics. Therefore, we conclude that an endemic state exists
whenever the solution $\rho_k=0$ is dynamically unstable. Following
this idea, we linearize Eq.~(\ref{eq:3}) around $\rho_k=0$, i.e.,
\begin{widetext}
\begin{equation}
  \frac{d \rho_k(t)}{dt} \approx-\bar{\delta}(k,\lambda) \rho_k(t)+
\lambda N \left[ \frac{k}{N \langle k
    \rangle}\right]^{\frac{b(\lambda)}{\ln{\kappa}}} 
  \sum_{k'} k'^{\frac{b(\lambda)}{\ln{\kappa}}}P(k') 
  \rho_{k'}(t). 
  \label{eq:4}
\end{equation}
\end{widetext}
This equation can be written in matrix form as
\begin{equation}
 \frac{d \rho_k(t)}{dt} \approx \sum_{k'} \mathbb{M}_{kk'}
 \rho_{k'}(t), 
\end{equation}
where
\begin{equation}
\mathbb{M}_{kk'}=-\bar{\delta}(k,\lambda) \delta_{kk'}+
\lambda N \left[ \frac{k}{N \langle k \rangle}\right]^{\frac{b(\lambda)}{\ln{\kappa}}}
   k'^{\frac{b(\lambda)}{\ln{\kappa}}}P(k').
\end{equation}
The stability of the absorbing state is then controlled by the maximum
eigenvalue of matrix $\mathbb{M}$, that is, the maximum $\Lambda_m$
solution of the eigenvalue problem $\mathbb{M} \vec{u}=\Lambda
\vec{u}$. In this way, $\Lambda_m=0$ defines the threshold between
the absorbing and endemic phases. The eigenvalue problem can be
rewritten as
\begin{equation}
\frac{\lambda N}{\Lambda+\bar{\delta}(k,\lambda)} \left[ \frac{k}{N
    \langle k \rangle}\right]^{\frac{b(\lambda)}{\ln{\kappa}}} 
  \sum_{k'} k'^{\frac{b(\lambda)}{\ln{\kappa}}}P(k') 
  u_{k'}=u_k,
\end{equation}
which gives us the explicit dependence of $u_k$ on $k$. Using this
result, the eigenvalues satisfy the equation
\begin{equation}
 \sum_{k} P(k) 
  \frac{\lambda N}{\Lambda+\bar{\delta}(k,\lambda)} \left[ \frac{k^2}{N \langle k \rangle}\right]^{\frac{b(\lambda)}{\ln{\kappa}}}=1.
\end{equation}
By setting $\Lambda=0$ and recalling that
$\tau(k,\lambda)=\bar{\delta}(k,\lambda)^{-1}$, we recover Eq.~(4) in
the main paper.

\section{Non small-world networks}

Let us assume that the average distance is given by
\begin{equation}
d=1+\alpha N^{\beta}.
\end{equation}
Plugging this expression in Eq.~(1) in the main paper, we obtain a
coarse description of the dynamics as
\begin{widetext}
\begin{equation}
  \frac{d \rho_k(t)}{dt}=-\bar{\delta}(k,\lambda) \rho_k(t)+
\lambda N e^{-\alpha N^{\beta}\frac{b(\lambda)}{\ln{\kappa}}}
  \sum_{k'} P(k') 
  \rho_{k'}(t)[1-\rho_k(t)]. 
\end{equation}
\end{widetext}
By repeating the same analysis performed in the previous section, we
conclude that the critical infection rate satisfies the equation
\begin{equation}
1=\lambda N e^{-\alpha N^{\beta}\frac{b(\lambda)}{\ln{\kappa}}}
  \sum_{k} P(k) \tau(k,\lambda).
\end{equation}
Assuming again that $P(k)$ decays slower than an exponential, this equation can be approximated as
\begin{equation}
1=\frac{\lambda}{a(\lambda)}
 e^{a(\lambda) k_{max}-\alpha N^{\beta}\frac{b(\lambda)}{\ln{\kappa}}-\ln\left[\frac{1}{NP(k_{max})}
   \right]}. 
\end{equation}
From this equation it is easy to see that for any fixed value of
$\lambda$, if $k_{max}$ grows faster than $N^{\beta}$, there exists a
size $N$ such that the exponent in this equation starts growing with
the system size and, therefore, the right hand side in this equation
will eventually grow above 1. The logical consequence is that, in this
case, the epidemic threshold goes to zero as $N$ goes to infinity.

\section{Robustness with respect to the coverage threshold $C$}

\begin{figure}[b]
\includegraphics*[width=\linewidth]{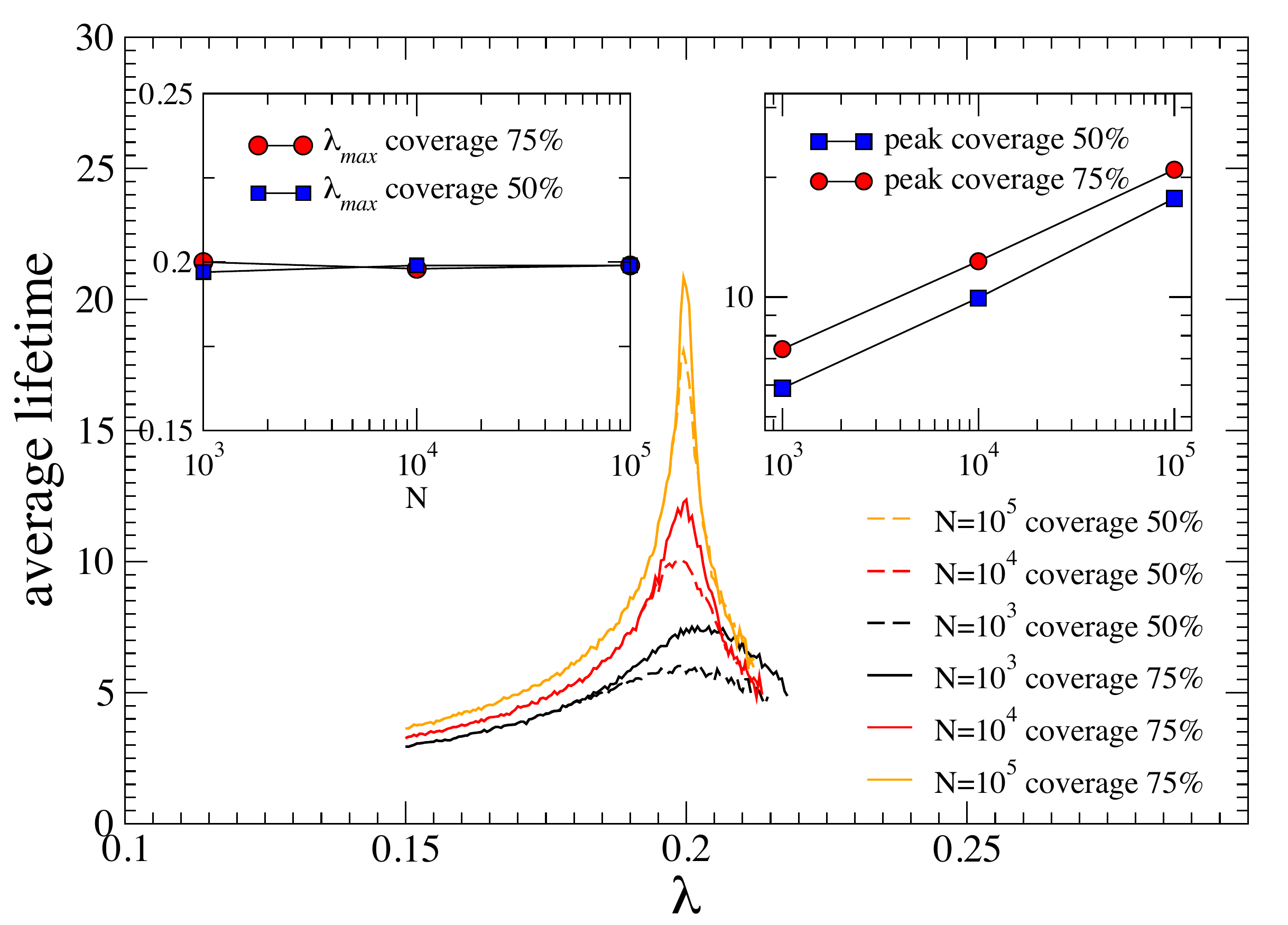}
\caption{Main: average lifetime $\bar{T}(\lambda,N)$ for finite realizations
as a function of $\lambda$ for the SIS model on an Erd\"os-R\'enyi graph
of average degree $\av{k}=5$. Left inset: position of the peak marking the
estimate of the numerical threshold as a function of the system size $N$.
Right inset: height of the peak as a function of $N$.}
\label{figS4} 
\end{figure}

\begin{figure}[t]
\includegraphics*[width=\linewidth]{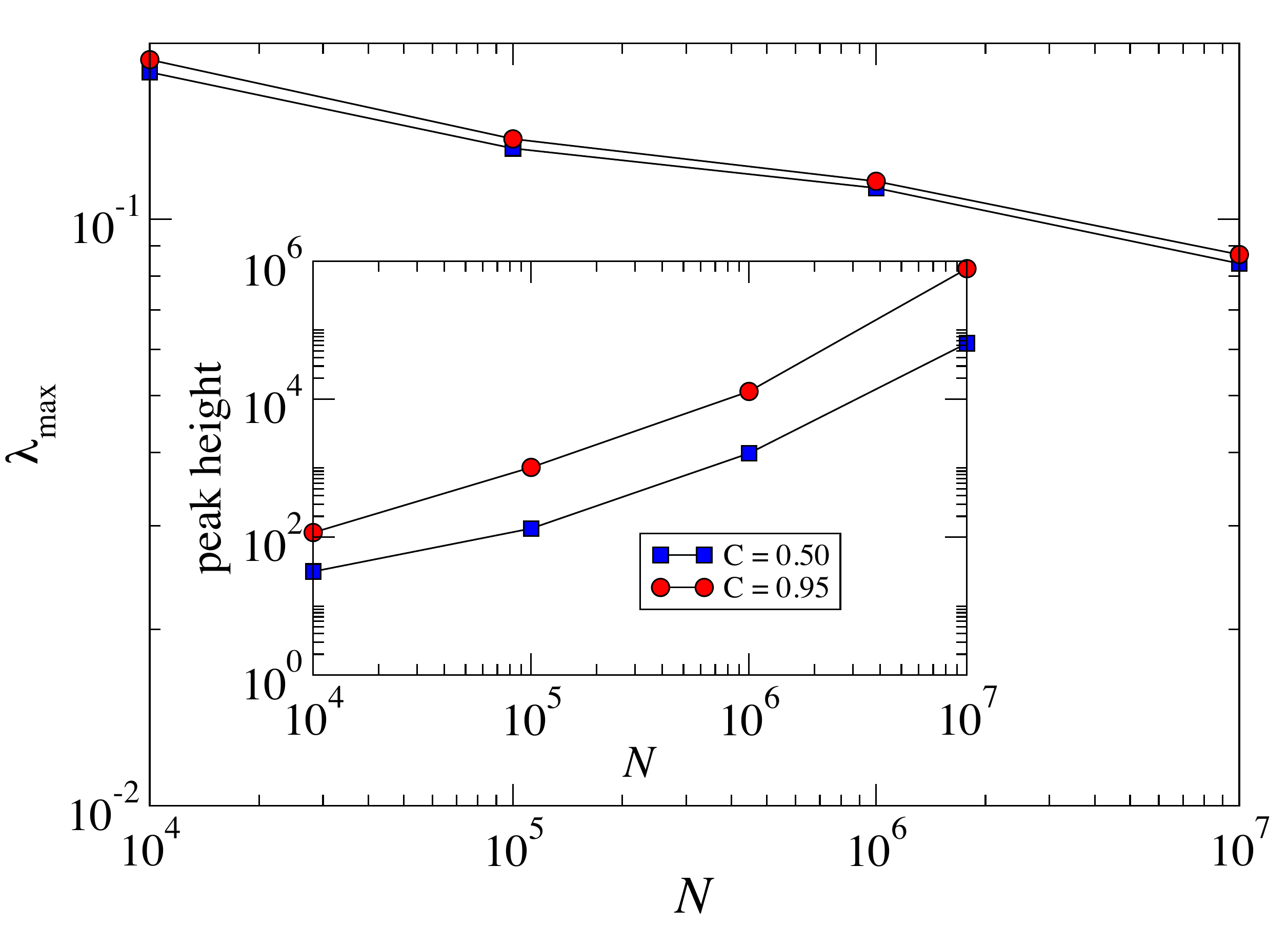}
\caption{Main: value of the epidemic threshold $\lambda_{max}$ estimated
numerically as the position of the peak of $\bar T(\lambda,N)$ as a 
function of $N$ for a UCM network with $\gamma=3.5$ and $k_{min}=3$.
Inset: height of the peak of $\bar T(\lambda,N)$ for the same system.}
\label{figS5} 
\end{figure}

To determine whether a given realization of the SIS process is
endemic, we have used the condition that the coverage is larger than a
fixed threshold value $C=0.5$. To check that different assumptions do
not qualitatively change the results, we performed some numerical
tests.  In Fig.~\ref{figS4}, we consider the effect of changing $C$
for an Erd\"os-R\'enyi graph of average degree $\av{k}=5$ and
different sizes $N$.  The numerical estimate of the threshold rapidly
converges to the expected value $\lambda_c=1/5$ for both values of $C$
considered, while the height of the peak grows with an exponent
independent of $C$.
In Fig.~\ref{figS5}, we perform the same analysis for a UCM graph with
$\gamma=3.5$ and $k_{min}=3$, obtaining similar results.  As the
system size $N$ is increased, the estimated thresholds decrease and
the peak heights increase in a perfectly analogous way.
Both figures confirm that the behavior of the model is robust with
respect to the arbitrary choice of the coverage threshold $C$.

\section{Further characterization of the epidemic transition}
\begin{figure*}[t]
\mbox{\includegraphics*[width=0.5\linewidth]{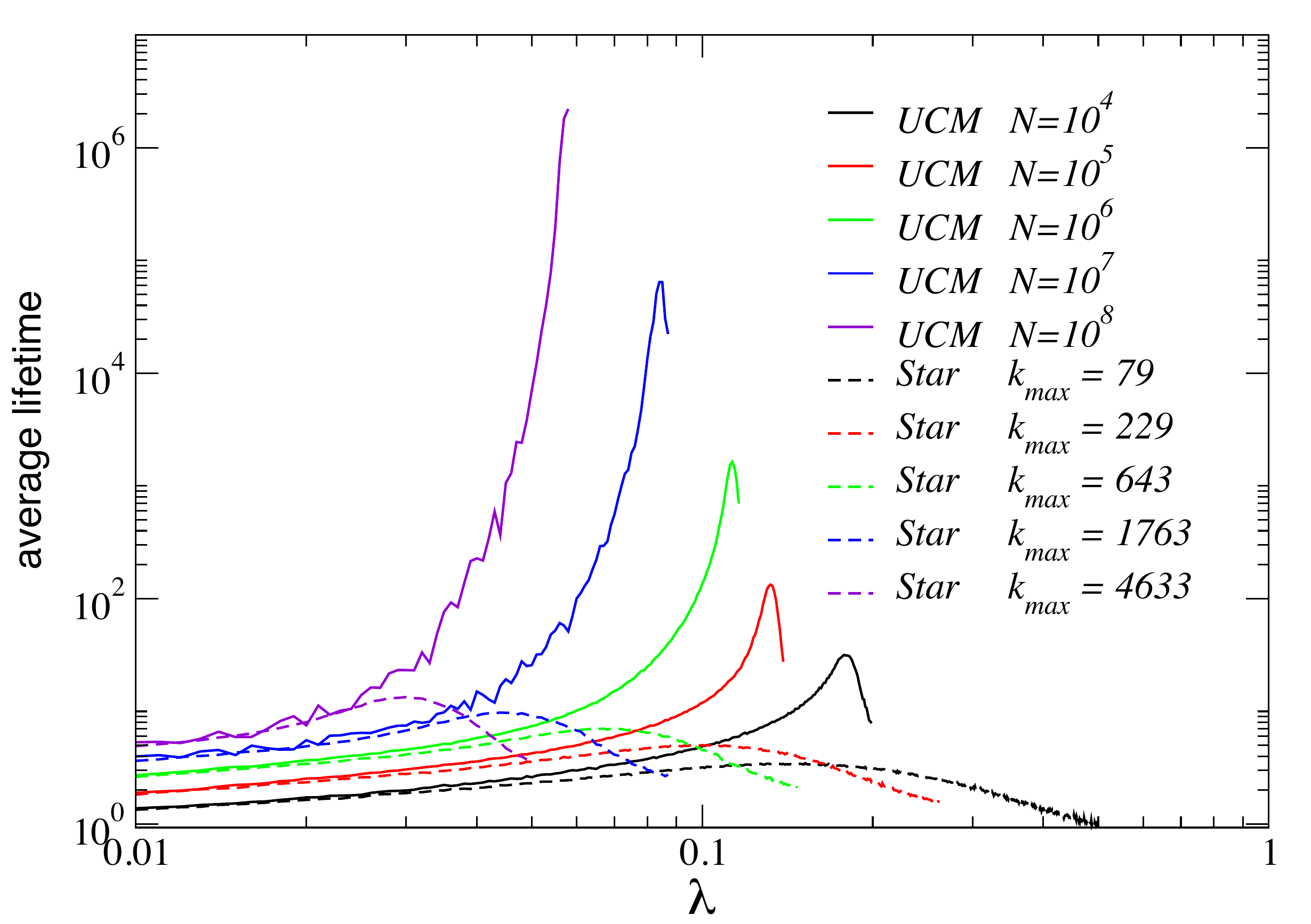}
\includegraphics*[width=0.5\linewidth]{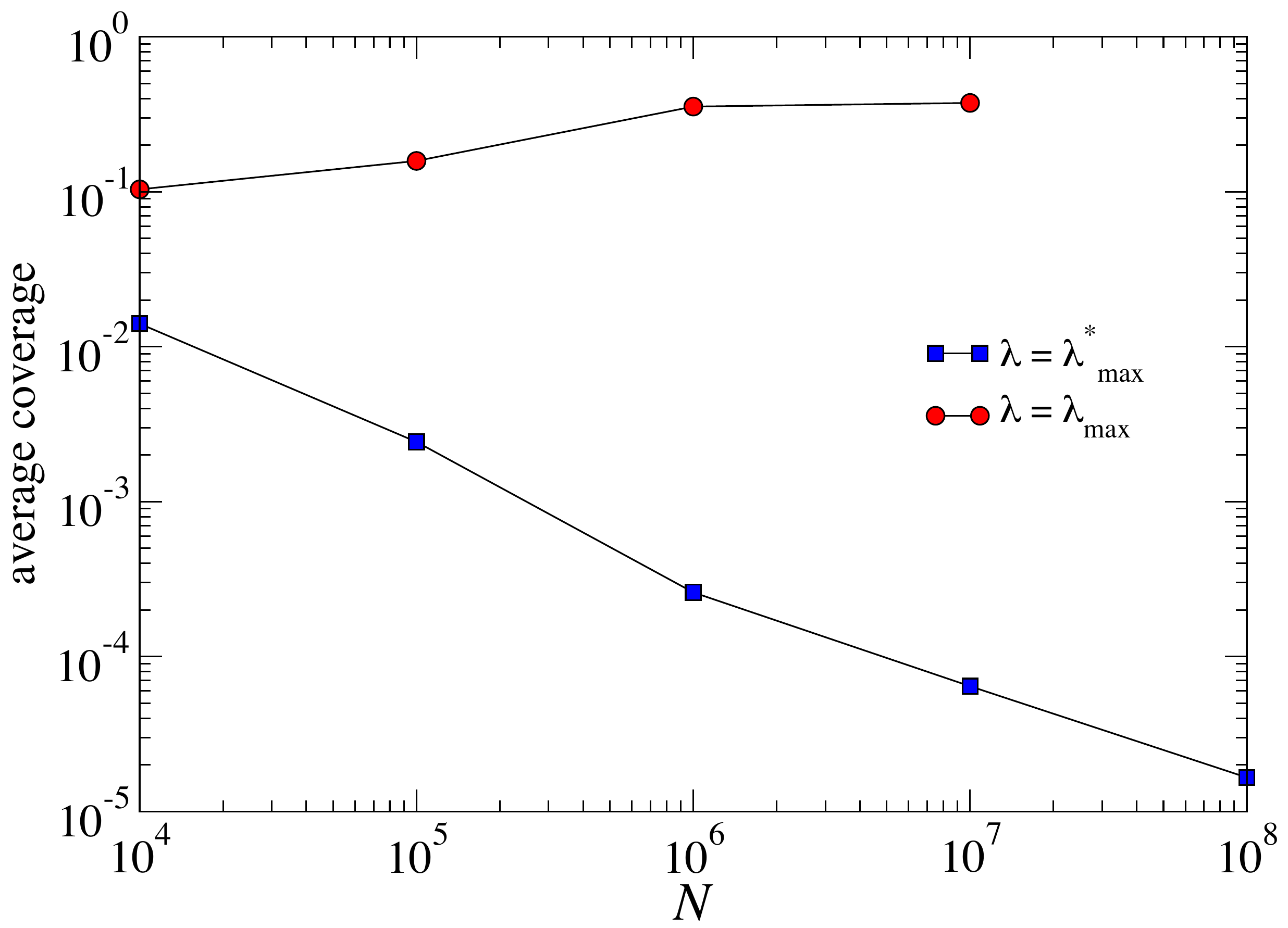}}
\caption{Left: Plot of the average lifetime of finite realizations
  $\bar T(\lambda,N)$ for UCM networks with $\gamma=3.5$ and
  $k_{min}=3$ and increasing values (bottom to top) of the system size
  $N$ (solid lines).  Dashed lines are the same quantity computed for
  a star graph made of $k_{max}+1$ nodes, where the values of
  $k_{max}$ are equal to the largest degree in the UCM
  networks. Right: Plot of the average coverage for a UCM network with
  $\gamma=3.5$ and $k_{min}=3$ for two different values of $\lambda$,
  corresponding to the thresholds for the whole network
  ($\lambda_{max}$) and for the star graph centered around the largest
  hub ($\lambda^*_{max}$). }
\label{figS6} 
\end{figure*}

In this section we provide some additional insight into the transition
marked by the peak of the average lifetime of finite realizations
$\bar T(\lambda,N)$.  In Fig.~\ref{figS6}, we compare the curves
already plotted in the top left of Fig.~1 of the main paper with the
analogous curves computed for a star graph with $k_{max}$ leaves.
It is clear that the occurrence of the peak in the full network is not
due only to the star graph centered around its hub.
The latter sustains alone the activity only for small $\lambda$.
The transition occurs at higher values of $\lambda$, for which the
lifetime is exceedingly larger.

In Fig.~\ref{figS6} we plot, as a function of $N$, the value of the
average coverage of the whole network at the critical value $\lambda_{max}(N)$
and at the critical value $\lambda_{max}^*(N)$ for the star graph centered around
the hub of degree $k_{max}$. 
It turns clearly out that at the critical point $\lambda_{max}$ the coverage
assumes a finite value in the thermodynamical limit, while it vanishes
for $\lambda=\lambda_{max}^*$. 
For $\lambda=\lambda_{max}^*$ the hub and its 
neighbors are fully covered, yet the epidemics does
not escape from the hub and its neighbors and it is thus localized.
At the transition point $\lambda_{max}$ instead the epidemics leaves the hub
and affects the whole population, leading to a truly endemic state.

\end{document}